\documentclass{aa}  

\usepackage{graphicx}
\usepackage{txfonts}
\makeatletter

\usepackage{graphicx}
\usepackage{booktabs}
\usepackage{placeins}
\usepackage{orcidlink}
\usepackage{tabularx}
\usepackage{hyperref}
\usepackage{comment}
\usepackage{caption}
\usepackage{footmisc}
\usepackage{footnote}
\usepackage[normalem]{ulem}
\usepackage{siunitx}
\begin{document}

   \title{Pulsar scintillation studies with LOFAR \\ \uppercase\expandafter{\romannumeral3}. 
   Annual variations in PSR~J0814$+$7429}

   \author{Yanqing Cai\orcidlink{0009-0009-9063-9118}\inst{1,2} 
               \and Ziwei Wu\orcidlink{0000-0002-1381-7859}\inst{1}
               \and Weiwei Zhu\orcidlink{0000-0001-5105-4058}\inst{1,3}
               \and Joris P. W. Verbiest\orcidlink{0000-0002-4088-896X} \inst{4}
               \and Yulan Liu\orcidlink{0000-0001-9986-9360}\inst{1,5,6}
               \and
                Krishnakumar Moochickal Ambalappat\inst{7}
                \and
                Marcus Brüggen\inst{8} \and  
                Benedetta Ciardi\inst{9} \and 
                Ralf-Jürgen Dettmar \orcidlink{0000-0001-8206-5956}\inst{10} \and 
               Ziyao Fang\inst{1}\and
               Qiuyang Fu\inst{1,2}\and
               Matthias Hoeft\orcidlink{0000-0001-5571-1369}\inst{11}\and
               Jiawei Jin\inst{1,2} \and
               Lars K\"unkel\orcidlink{0000-0003-4634-5453}\inst{12,13}\and
               J\"orn K\"unsemöller \inst{4}\and
               Caisong Liu\inst{1,2}\and
               Lingqi Meng\orcidlink{0000-0002-2885-568X}\inst{1,2}\and
               Xueli Miao\orcidlink{0000-0003-1185-8937}\inst{1}\and
               Jiarui Niu\orcidlink{0000-0001-8065-4191}\inst{1}\and
               Rukiya Rejep\orcidlink{0000-0002-3283-075X}\inst{1}\and
               Dominik J. Schwarz\orcidlink{0000-0003-2413-0881}\inst{4} \and 
               Golam M. Shaifullah\orcidlink{0000-0002-8452-4834}\inst{14,15}\and
               Caterina Tiburzi\orcidlink{0000-0001-6651-4811}\inst{16} \and 
               Christian Vocks \orcidlink{0000-0001-8583-8619}\inst{17} \and
               Olaf Wucknitz \orcidlink{0000-0001-8348-4861}\inst{18} \and     
               Mengyao Xue\orcidlink{0000-0001-8018-1830}\inst{1}\and
               Mao Yuan\orcidlink{0000-0003-1874-0800}\inst{19}\and
               Youling Yue\orcidlink{0000-0003-4415-2148}\inst{1}\and
               Chunfeng Zhang\orcidlink{0000-0002-4327-711X}\inst{1}\and 
               Zhen Zhang\inst{3}
               }

   \institute{State Key Laboratory of Radio Astronomy and Technology, National Astronomical Observatories, Chinese Academy of Sciences, Beijing 100101, China\\
            \email{wuzw@bao.ac.cn,zhuww@nao.ac.cn}
              \and
              University of Chinese Academy of Sciences,
             Chinese Academy of Sciences, Beijing 100049, China
             \and 
             Institute for Frontier in Astronomy and Astrophysics, Beijing Normal University, Beijing 102206, China 
             \and
             Fakult\"at f\"ur Physik, Universit\"at Bielefeld, Postfach 100131, 33501 Bielefeld, Germany
             \and
             CAS Key Laboratory of FAST, National Astronomical Observatories, Chinese Academy of Sciences, Beijing 100101, China
             \and 
             Guizhou Radio Astronomical Observatory, Guizhou University, Guiyang 550001, China
             \and 
            National Centre for Radio Astrophysics, Tata Institute of Fundamental Research, Pune 411007, Maharashtra, India
            \and
            Hamburger Sternwarte, University of Hamburg, Gojenbergsweg 112, 21029 Hamburg, Germany
            \and
            Max Planck Institute for Astrophysics, Karl-Schwarzschild-Str 1, D-85741 Garching, Germany
            \and
            Ruhr-Universität Bochum, Fakultät für Physik und Astronomie, Astronomisches Institut, 44780 Bochum, Germany 
            \and
            Th\"uringer Landessternwarte Tautenburg,  Sternwarte 5, 07778 Tautenburg, Germany
            \and
            Department of Physics, McGill University, 3600 rue University, Montr\'eal, QC H3A 2T8, Canada
            \and
            Trottier Space Institute, McGill University, 3550 rue University, Montr\'eal, QC H3A 2A7, Canada
            \and
            Dipartimento di Fisica ``G. Occhialini'', Universit\`a di Milano-Bicocca, Piazza della Scienza 3, 20126 Milano, Italy 
            \and
            INFN, Sezione di Milano-Bicocca, Piazza della Scienza 3, I-20126 Milano, Italy 
            \and
            INAF - Osservatorio Astronomico di Cagliari, via della Scienza 5, 09047 Selargius (CA), Italy
            \and
            Leibniz-Institut f\"ur Astrophysik Potsdam (AIP), An der Sternwarte 16, 14482 Potsdam, Germany
            \and
            Max-Planck-Institut f\"ur Radioastronomie, Auf dem H\"ugel 69, 53121 Bonn, Germany
            \and
            National Space Science Center, Chinese Academy of Sciences, Beijing 100190, China             
            }
            
   \date{Received XXX; accepted XXX}
 
  \abstract
   {The interstellar scintillation observed in radio pulsars arises from interference between electromagnetic waves scattered by electron density fluctuations in the turbulent interstellar plasma, providing a critical tool for probing the small-scale structure of the ionized interstellar medium and the pulsar system itself.
   }
   {The primary aim of this work is to study long-term scintillation variations for a bright and nearby pulsar, PSR J0814$+$7429, carried out from 2013 September to 2023 September with the LOw-Frequency ARray (LOFAR) High Band Antennae in the frequency range of 120 - 170 MHz. }
   {We derive the basic scintillation parameters, scintillation bandwidth ($\Delta\nu_{\rm d}$) and scintillation timescale ($\Delta\tau_{\rm d}$), from the two-dimensional (2D) auto-covariance function of the dynamic spectra that are a 2D matrix of pulse intensity as a function of time and frequency.}
   {We present the long-term monitoring of $\Delta\nu_{\rm d}$ and $\Delta\tau_{\rm d}$ for PSR J0814$+7429$, which shows a strong annual variation in the time series of the $\Delta\tau_{\rm d}$. From our modeling of the annual variations of scintillation velocities, the scattering screen is anisotropic and located at $0.23\pm0.02$ kpc from the Earth, likely corresponding to the boundary of the Local Bubble.
    }
   {}

   \keywords{pulsar:individual:PSRJ0814+7429 --
                ISM: general
               }

   \maketitle
%

\section{Introduction}
Pulsars are highly magnetized, rapidly rotating neutron stars that appear to emit highly periodic electromagnetic radiation pulses as their lighthouse-like radiation beams sweep across the observer.
Pulsar emission can be detected across various wavelengths such as radio \citep{hbp+68}, X-ray \citep{bhw+14}, and gamma rays \citep{aaa+13}. 
As these radio signals of pulsars propagate through electron density turbulence in the ionized interstellar medium (IISM), they are scattered, and then the interference between these scattered signals results in a modulation of the pulse intensity as a function of frequency and time, forming an interference pattern on the observer plane \citep{ric90}. 
This phenomenon, known as interstellar scintillation \citep[ISS,][]{sch68}, has become an important tool for probing the properties of the IISM \citep{ric77,ric90,nar92a} and studying binary pulsar systems \citep{lyn84,rcn+14}. 
In addition, analysis of ISS in precisely timed pulsars may be used to generate timing corrections for gravitational-wave characterization by pulsar timing arrays \citep{vob21}, such as the Chinese Pulsar Timing Array \citep{cpta23}, the European Pulsar Timing Array \citep{epta23}, the Parkes Pulsar Timing Array \citep{ppta23}, and the North-American Nanohertz Observatory for Gravitational Waves \citep{usa23}, then enhancing the probability of detecting gravitational waves.

There are two main types of ISS: diffractive ISS (DISS) and refractive ISS (RISS).
DISS arises from small-scale density fluctuations in the IISM and is typically observed with characteristic timescales ranging from minutes to hours \citep{rik69,wvm+22, lvm2022}.
RISS originates from larger-scale density gradients and manifests over weeks to months \citep{sie82, rl90}.
The interference effects produced by ISS are measurable from dynamic spectra, which are two-dimensional representations of pulse intensity as a function of observing time and radio frequency.
The characteristic time and frequency widths of scintles can be parameterized using the scintillation timescale $\Delta \tau_{\rm d}$ and scintillation bandwidth $\Delta \nu_{\rm d}$, which both can be obtained from the two-dimensional autocorrelation function (2D ACF) of the dynamic spectrum \citep{cpl86,wmj+05}. 
In general, $\Delta \nu_{\rm d}$ is highly frequency dependent and $\Delta \tau_{\rm d}$ is relatively weakly frequency dependent, and also modulated by the relative motions between the pulsar, the scattering medium, and the observer. 

In some cases, the long time series of $\Delta\tau_{\rm d}$ shows an annual variation, and it is also possible to show orbital variations if the pulsar is in a binary system \citep{lyn84}. 
By employing these variations, we can determine the small-scale distribution and inhomogeneities of the IISM, or some orbital parameters which are difficult to measure with the pulsar timing method alone. 
For example, \cite{rch+19} used long-term scintillation of PSR J1141$-$6545, which shows orbital and annual variations, to resolve ambiguity in the sense of the inclination angle. 
Recently, \cite{lmv+23} also used the annual and orbit of variations to measure the distance, velocity, and degree of anisotropy of the scattering screen for PSRs J0613$-$0200 and J0636$+$5128, their results further add to the growing evidence of the Local Bubble shell as a dominant region of scattering along many lines of sights. 
The Local Bubble, which is a cavity of hot, tenuous, X-ray emitting gas surrounding the solar system, likely created by past supernova explosions or stellar winds \citep{cr87,ms+90}.
In addition to the above pulsars, the following pulsars also had their annual and orbital variations studied: J0737$-$3039A \citep{rcn+14}, J0437$-$4715 \citep{rcb+20}, and J1603$-$7202 \citep{wrts22}.
However, the number of pulsars that have been reported with orbit or annual variations is very small, accounting for only a small fraction of the total number of pulsars. 

The LOw Frequency ARray \citep[LOFAR,][]{vwg+13} provides unique advantages for ISS studies, owing to the strong frequency dependence of ISS. 
Previous works have already presented the census of DISS pulsars at LOFAR frequencies \citep{wvm+22} and also reported on non-standard frequency-dependence of the scattering properties using near-simultaneous observations from LOFAR and the New extension in Nan\c{c}ay upgrading LOFAR \citep[NenuFAR,][]{wcv+23}.
In this paper, we present long-term scintillation monitoring of PSR J0814$+$7429 with LOFAR. 
This pulsar was discovered by \cite{cp68}, and it provided one of the richest contexts for studying pulsar emission physics. 
In Section \ref{sec2}, we describe our observations and data processing. 
The analysis results and the discussion are presented in Section \ref {sec3}. 
The conclusion is given in Section \ref {sec4}.

\begin{figure}
    \centering
    \includegraphics[width=1\linewidth]{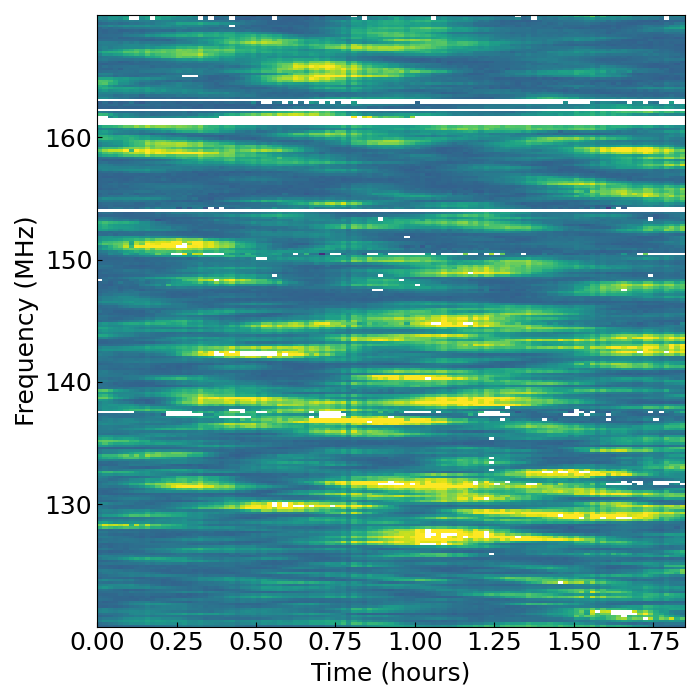}
    \caption{Example dynamic spectrum of PSR J0814$+$7429 with LOFAR taken on MJD 57361. The white patches are removed due to RFI.}
    \label{fig:dynamic}
\end{figure}

\begin{table}
    \centering
    \caption{LOFAR HBAs observational campaign for PSR J0814$+$7429.}
    \begin{tabular}{ccS[table-format=3.0, table-number-alignment=center] S[table-format=4.2, table-number-alignment=center]}
    \hline
    \hline
    Station & Date range & \text{Number of obs.} & \text{Length}\\
    & (year) & & \text{(hours)}\\
    \hline
    DE601 & 2013$-$2018 & 70 & 94.58 \\
    DE602 & 2016$-$2017 & 18 & 43.26 \\
    DE603 & 2014$-$2020 & 42 & 46.20 \\
    DE604 & 2016$-$2017 & 58 & 139.95\\
    DE605 & 2014$-$2019 & 254 & 436.82 \\
    DE609 & 2017$-$2023 & 862 & 1827.12 \\
    \hline
    \multicolumn{4}{p{\linewidth}}{\small {\bf Notes:} Given are the LOFAR stations used and their observation date range, along with the number and the total length of observations for each station.}
    \end{tabular}
    
    \label{label_station}
\end{table}

\section{Observations and data processing}\label{sec2}

\begin{figure*}[h]
    \centering
    \includegraphics[width=1\textwidth]{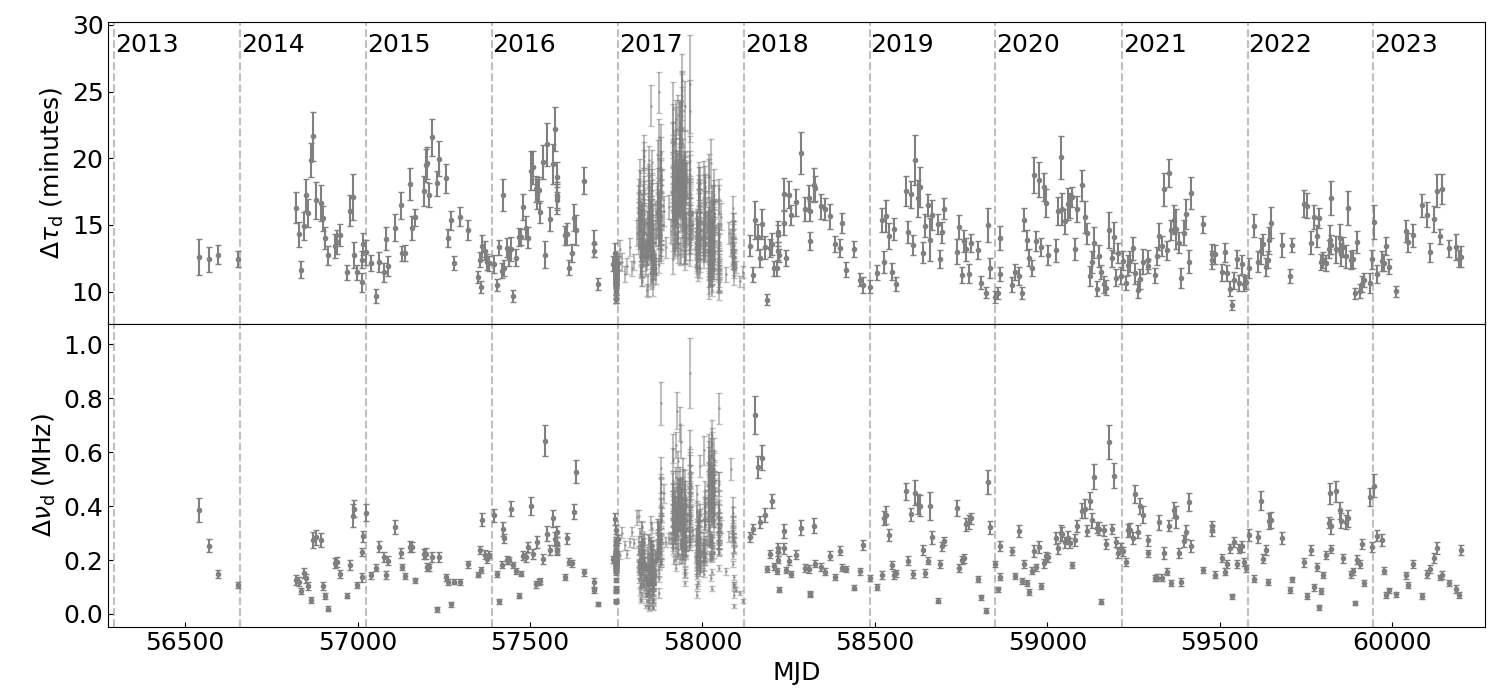}
    \caption{Long-term variations of scintillation timescale $\Delta \tau_{\rm d}$ (top panel) and scintillation bandwidth $\Delta \nu_{\rm d}$ (bottom panel) for PSR J0814$+$7429. The vertical gray dotted lines indicate the start of a calendar year. More observations in 2017 resulted in relatively dense data points this year.}
    \label{fig:oneD_time_freq}
\end{figure*}

The observations of PSR~J0814+7429 presented in this study were conducted between September 2013 and September 2023, covering a selected frequency range of 120–170~MHz after omitting the first and last 10 MHz bands contaminated by radio frequency interference (RFI). 
Over this decade-long campaign, we collected 1304 high band antenna (HBA) observations. 
Six LOFAR stations were utilized for these measurements: DE601-Effelsberg, DE602-Unterweilenbach, DE603-Tautenburg, DE604-Bornim, DE605-J\"ulich, and DE609-Norderstedt (see Table \ref{label_station} for specifications).
All observations used in our work have been done in stand alone mode using the stations individually.
To ensure coverage of more scintles in the dynamic spectra, our analysis focuses on observations lasting at least 30 minutes. 
The data are recorded with a frequency resolution of 0.195 MHz per channel and a typical temporal resolution of 60 seconds per subintegration. 
The data processing pipeline closely follows the methodology described by \cite{wvm+22}; further details on observation protocols, raw data reduction, and scintillation parameter extraction can be found in Section 2 of that work.
After data processing, all dynamic spectra are visually inspected, and those seriously contaminated by RFI are excluded using \texttt{ITERATIVE$\_$CLEANER}\footnote{\url{https://github.com/larskuenkel/iterative_cleaner}.}.
A representative cleaned dynamic spectrum is displayed in Figure \ref{fig:dynamic}.
The long-term evolution of the scintillation bandwidth ($\Delta \nu_{\rm d}$, bottom panel) and timescale ($\Delta \tau_{\rm d}$, top panel) for PSR~J0814+7429 is presented in Figure \ref{fig:oneD_time_freq}. 
Notably, the $\Delta \tau_{\rm d}$ time series exhibits a pronounced annual variation, which provides critical insights into the properties of the IISM, as discussed in the next Section.

\section{Results and Discussion} \label{sec3}
In this section, we present our methodology for modeling the annual variations in scintillation properties observed in PSR J0814+7429 and report the corresponding analysis results. 
For a more comprehensive analysis of annual scintillation variations and detailed discussions of these phenomena in pulsar systems, we direct readers to the recent work by \citealt{rch+19, lmv+23}.

\subsection{Scintillation from a thin screen} 
\label{sec3.1}

The distribution of IISM responsible for the observed ISS is complex along the line of sight between the pulsar and the observer. 
In this paper, we consider a simple thin screen model for the medium distribution in the case of PSR~J0814+7429, which is supported by the detection of a scintillation arc \citep{wvm+22}.
For a thin screen model, \cite{cr98} presented that the scintillation velocity $V_{\rm ISS}$, which is the ratio of the spatial scale of the diffraction pattern to the scintillation timescale $\Delta \tau_{\rm d}$, can be estimated from the scintillation parameters,
\begin{equation}\label{exp1}
 V_{\rm ISS}=A_{\rm ISS} \frac{\sqrt{D\Delta \nu_{\rm d}}}{f\Delta \tau_{\rm d}}, 
\end{equation}
where $D$ is the distance from the observer to the pulsar in kpc, $f$ is the observing frequency in GHz, $\Delta \nu_{\rm d}$ is in units of MHz and $\Delta \tau_{\rm d}$ is in units of seconds. 
The factor $A_{\rm ISS}$ depends on the geometry, the location of the scattering screen, and the form of the turbulence spectrum. 
\cite{cr98} derived $A_{\rm ISS}=2.5\times 10^{4} $ km s$^{-1}$ for a uniform medium, while for a thin screen model, $A_{\rm ISS}=2.78\times 10^{4}\sqrt{2D_{\rm s}/(D-D_{\rm s})}$ km s$^{-1}$, where $D_{\rm s}$ represents the distance of the scattering screen from the Earth and is in units of kpc. 
Over the past two decades, there has been increasing evidence of anisotropic scattering in many cases, including pulsars \citep{smc+01,bmg+10, wvm+22, mpj+23} and fast radio bursts \citep{wmz+24,wzz+24}. 
Moreover, \cite{srm+22} presented that approximately 20\% of the pulsars within their sample show evidence of the existence of anisotropy of scattering.  
Therefore, in this work we consider an anisotropic scattering screen and the anisotropy dependent $A_{\rm ISS}$, which can be given by $A_{\rm ISS}=2.78\times 10^{4} \sqrt{(A_{\rm r}+1/A_{\rm r})/2}\sqrt{2D_{\rm s}/(D-D_{\rm s})}$ \citep{rcn+14}, where $A_{\rm r}$ is the axial ratio of anisotropy under assuming the spatial diffraction pattern as an ellipse \citep{cmr+05}.

In the thin screen model, the scintillation velocity $V_{\rm ISS}$ at the observer can be predicted as \citep{lmv+23}:
\begin{equation}\label{exp2}
    V_{\rm ISS}=|V_{\rm eff}|\frac{D}{D-D_{\rm s}} ,
\end{equation}
where $V_{\rm eff}$ represents the effective scintillation velocity which is a combination of the pulsar's, the Earth's and the IISM's transverse velocities weighted by the fraction of distance of the scattering screen as \citep{cr98}:
\begin{equation}\label{exp3}
    V_{\rm eff}=\frac{D-D_{\rm s}}{D}V_{\rm E}+\frac{D_{\rm s}}{D}(V_{\rm p}+V_{\rm \mu})-V_{\rm IISM},
\end{equation}
where $V_{\rm E}$ is the Earth velocity, $V_{\rm p}$ is the pulsar's binary orbital transverse velocity, $V_{\rm \mu}$ is the pulsar proper motion transverse velocity and $V_{\rm IISM}$ is the transverse velocity of the IISM. 
For a solitary pulsar system, $V_{\rm p}$ is zero. 
Similar with \citealt{lmv+23}, we integrate Equation \ref{exp1} and Equation \ref{exp2} into the following formula,
\begin{equation}\label{exp4}
    \frac{\sqrt{\Delta\nu_{d}}}{f\Delta\tau_{d}}\equiv Q =\frac{|V_{\rm eff}|}{A_{\rm ISS}}\frac{\sqrt{D}}{D - D_{\rm s}},
\end{equation}
where all the scintillation observables are combined in the definition of $Q$.
The uncertainty of $Q$ is mainly from the statistical uncertainties of scintillation parameters.
\begin{figure*}
    \centering
    \includegraphics[width=0.8\linewidth]{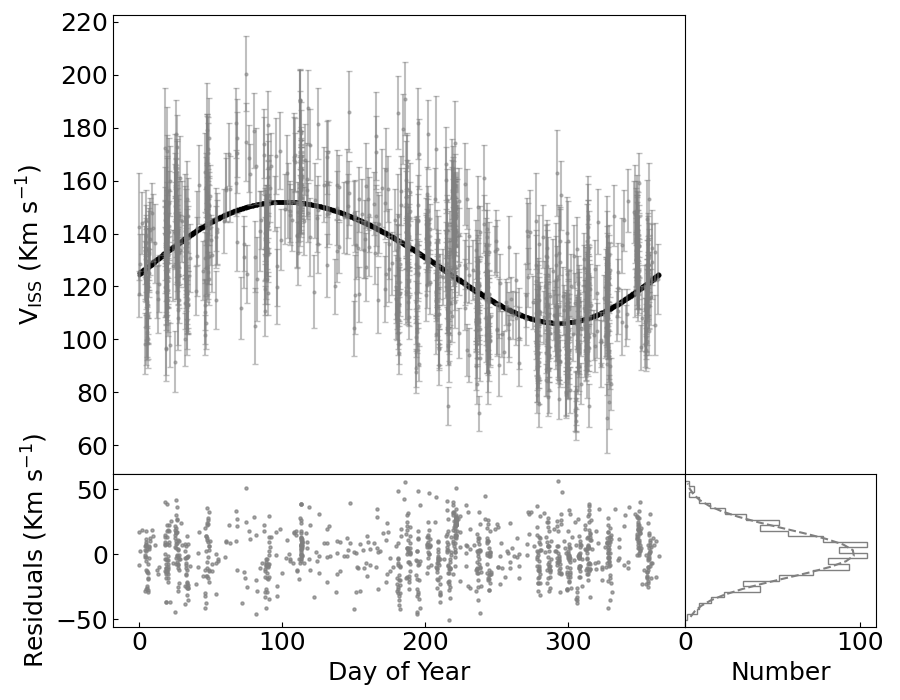}
    \caption{Top panel: the scintillation velocities of PSR J0814$+$7429 as a function of day of year. The points with error bars are the scintillation velocities from Equation \ref{exp1}, and the curve shown is that from Equation \ref{exp2}. The residuals between the two scintillation velocities are shown in the bottom left panel. And the residual histogram distribution with Gaussian fitting (gray dashed line) is shown in the bottom right panel.}
    \label{fig:viss_annual}
\end{figure*}

For the effective transverse velocity $|V_{\rm eff}|$, we take it in equatorial coordinates employing the quadratic form from \cite{rcn+14} as follows:
\begin{equation}
\begin{split}
    |V_{\rm eff}| &= \sqrt{ aV_{\rm eff, \alpha}^{2}+bV_{\rm eff, \delta}^{2}+cV_{\rm eff, \alpha}V_{\rm eff, \delta}},\\
    V_{\rm eff, \alpha} &= \frac{D-D_{\rm s}}{D}V_{\rm E, \alpha}+\frac{D_{\rm s}}{D}V_{\mu, \alpha}-V_{\rm IISM, \alpha},\\
    V_{\rm{eff}, \delta} &= \frac{D-D_{\rm s}}{D}V_{\rm E, \delta}+\frac{D_{\rm s}}{D}V_{\mu, \delta}-V_{\rm IISM, \delta},
\end{split}  
\end{equation}
where $V_{\rm eff, \alpha}$ and $V_{\rm eff, \delta}$ are the components of the effective velocity in right ascension ($\alpha$) and in declination ($\delta$), respectively. 
As the assumption that the spatial diffraction pattern is an ellipse mentioned above, the coefficients $a$, $b$, and $c$ are parameterized by the axial ratio $A_{\rm r}$ of the ellipse and the orientation angle $\psi$ of the $A_{\rm r}$. 
Considering the orientation angle $\psi$ is defined clockwise from the right ascension of the pulsar, the coefficients are \citep{rcn+14}:
\begin{equation}
    \begin{split}
        a &= [1-R\cos(2\psi)]/\sqrt{1-R^{2}},\\ 
        b &= [1+R\cos(2\psi)]/\sqrt{1-R^{2}},\\
        c &= -2R\sin(2\psi)/\sqrt{1-R^{2}},
    \end{split}
\end{equation}
where $R$ is a scaled parameter for $A_{\rm r}$: $R = (A_{\rm r}^{2}-1)/(A_{\rm r}^{2}+1)$, which is bound between 0 and 1.

\subsection{Fitting annual variations}
Similar to \cite{rch+19} and \cite{lmv+23}, for the value of the $\Delta\nu_{\rm d}$ in Equation \ref{exp1} we use the average value of all $\Delta\nu_{\rm d}$ observations. 
In Equation \ref{exp3}, $V_{\rm eff}$ incorporates three transverse velocity components for PSR J0814$+$7429: the pulsar’s proper motion, Earth’s orbital motion, and the velocity of the scattering screen. 
We derive Earth’s velocity using SCINTOOLS\footnote{\url{https://github.com/danielreardon/scintools/}} \citep{rcb+20}.
The pulsar’s transverse velocity, $V_{\mu}$, is obtained from \cite{bbgt02}. 
The distance to PSR~J0814$+$7429, D=0.432 kpc, is adopted from \cite{vwc+12}. 
Finally, the scintillation velocity model of PSR J0814$+$7429 includes five free parameters in the anisotropic scattering case: $V_{\rm IISM, \alpha}$, $V_{\rm IISM, \delta}$, $A_{\rm r}$, $\psi$ and $D_{\rm s}$.

We model the time series of the parameters $Q$ using Markov Chain Monte Carlo (MCMC) with EMCEE \citep{fhlg13}. 
All fitting parameters have uniform priors, and their posterior distributions are presented in Fig. \ref{fig:enter-mcmc}.
We computed a reduced $\chi^2$ value of 2.40 for the thin screen model. 
This value is slightly larger than unity, but the residuals between the observed and predicted scintillation velocities in the bottom panel of Figure \ref{fig:viss_annual} are approximately Gaussian-distributed and exhibit no systematic patterns, indicating that the model provides an overall satisfactory description of the data. 
The modest deviation from unity likely reflects minor underestimation of the measurement uncertainties due to the limited frequency resolution, rather than a significant deficiency in the model itself.
Therefore, considering possible underestimation of measurement uncertainties, this model can provide a reasonably good fit to the data.
The best-fit parameters for $V_{\rm IISM, \alpha}$, $V_{\rm IISM, \delta}$, $A_{\rm r}$, $\psi$ and $D_{\rm s}$ are $-50\pm10$ km s$^{-1}$, $-21^{+9}_{-8}$ km s$^{-1}$, $2.0\pm0.3$, $168^{+7}_{-8}$ degrees and $0.23\pm0.02$ kpc, respectively.
The axial ratio $A_{\rm r}$ of the spatial diffraction pattern is $2.0\pm0.3$, which is in agreement with the mildly anisotropic scattering model. 
Our results would predict a scintillation arc curvature of $\sim$4.2~$s^{3}$ at 145~MHz at MJD 57872, which is different from 2.5~$s^{3}$ \citep{wvm+22} using $\eta \propto f^{-2}$ \citep{sro+19}.  
This may indicate that there are multiple scattering screens along the line of sight \citep{occ+24}. 
Fig. \ref{fig:viss_annual} shows the variation of the scintillation velocities as a function of day of year, where the scintillation velocity from Equation \ref{exp1} is denoted by the gray point with error bar, and that from Equation \ref{exp2} by the solid black curve. 
One can see that there is a clear annual variation in these scintillation velocities, caused by the annual change in the Earth's velocity.

The values of $|V_{\rm IISM, \alpha}|$ and $|V_{\rm IISM, \delta}|$ seem to be larger than the expected average plasma velocity of about 10 km s$^{-1}$ \citep{gs95} in the IISM. 
After considering the differential rotation of the galaxy, the transverse velocity of the scattering screen is still about 48~km/s. 
Similar larger IISM velocities in other pulsars are also reported \citep{rcn+14}.
Due to the limited understanding of the plasma's origin, it is difficult to determine whether these larger IISM velocities are unusual or unreasonable \citep{rcb+20}.

The screen distance is measured to be $D_{\rm s}=0.23\pm0.02$ kpc for PSR J0814$+$7429.
\cite{rcm00} studied the characteristics of the scattering medium through long-term monitoring of the weak scintillation for PSR J0814$+$7429.
And they proposed that scattering of this pulsar is consistent with either an extended scattering medium along the entire line of sight or a localized compact structure in the range 170$-$220 pc (with a pulsar distance of 310 pc \citep{tc93}) from the Earth.
Their distance-revised screen distance is fully consistent with our result.

Currently, observed ISS phenomena have been suggested to be caused by the boundary of the Local Bubble \citep[e.g.,][]{bgr98}, H\MakeUppercase{\romannumeral 2} regions \citep{mma+22}, hot stars \citep{wtb+17}, and supernova remnants \citep{yzm+21}.
We find no associated H\MakeUppercase{\romannumeral 2} regions in a search of the Wide-field Infrared Survey Explorer \citep[WISE:][]{abb14} along the line of sight for PSR~J0814$+$7429.
We further search the \textit{Gaia} DR3 catalogue \citep{gvb+23} for the presence of hot stars in close proximity to the line of sight, without reliable hot stars to contribute to the scattering of PSR J0814$+$7429. 
But the distance of the scattering screen of PSR J0814$+$7429 is consistent with the shell of the local bubble, which is similar to earlier findings for PSRs J0613$-$0200 and J0636$+$5128 \citep{lmv+23}, implying that the evidence of the boundary of the local bubble may be a dominant region of scattering along the line of sight to PSR J0814$+$7429.
In addition, recent low-frequency Faraday tomography observations from the LOFAR Two-Metre Sky Survey \citep[LoTSS,][]{ejh+24,ejh24} revealed that the Local Bubble is bounded by a magnetised and partially ionised shell located about 40–200 pc, which potentially contributes to the observed scintillation behaviours of pulsars.
However, recent work by \cite{occ+24} mentioned that accurately associating the scattering screens with the Bubble's boundary remains difficult, due to uncertainties in both the screen distances and the Bubble's modeled surface. More observations are needed to further provide additional evidence linking scattering screens of pulsars to the Bubble's boundary.

\section{Conclusion} \label{sec4}
In this paper, we have presented an analysis of long-term scintillation monitoring for the solitary PSR J0814$+$7429. 
This has been accomplished across a data span of $\approx$10 yr observed with LOFAR. 
The annual variation in ISS can be explained as an annual modulation by the orbital motion of the Earth. 
Under the framework of an anisotropic thin scattering screen model, we have modeled the annual variations of scintillation parameters. 
The value of axis ratio $A_{\rm r}$ is $2.0\pm0.3$, which is in agreement with the mildly anisotropic scattering model. 
We also constrain the screen distance approximately halfway between the pulsar and the Earth at $D_{s}=0.23\pm0.02$ kpc. 
Along the line of sight at this distance, we suggest that the scattering screen of PSR J0814$+$7429 is very likely associated with the boundary of the Local Bubble.

For pulsars with scattering screens close to Earth,
long-term scintillation monitoring for them could more readily reveal annual variations similar to those observed in PSR J0814+7429.
We have once again demonstrated the feasibility of using LOFAR for ISS studies of pulsars.
More long-term ISS monitoring on other pulsars will be released, hoping to provide more clues to identify the ISS mechanism and the astrophysical structures responsible for ISS.

\begin{acknowledgements}
We thank the anonymous referee for the constructive comments and suggestions, which helped us to improve the presentation of this  paper.
This work is supported by NSFC grant No.~12503056, grants from Beijing Nova Program  (No.~20250484786), National Science Foundation, China, no. 12421003, the CAS-MPG LEGACY project, the Max-Planck Partner Group, the Strategic Priority Research Program of the Chinese Academy of Sciences (Nos. XDA0350501), and the Major Science and Technology Program of Xinjiang Uygur Autonomous Region, grant No. 2022A03013-2.
J.P.W.V.\ acknowledges support by the Deutsche Forschungsgemeinschaft (DFG) through the Heisenberg programme (Project No.\ 433075039).
MB acknowledges support by the Deutsche Forschungsgemeinschaft (DFG, German Research Foundation) under Germany’s Excellence Strategy – EXC 2121 `Quantum Universe' – 390833306.
MH acknowledges support from the Federal Ministry of Research, Technology and Space (BMFTR) ErUM-Pro under grant 05A23STA.
RJD acknowledges support from BMFTR ErUM-Pro under grant 05A23PC2. DJS acknowledges support from BMBFT ErUM-Pro under grant 05A23PB1.
LOFAR \citep{vwg+13} is the Low Frequency Array designed and constructed by ASTRON. It has observing, data processing, and data storage facilities in several countries, that are owned by various parties (each with their own funding sources), and that are collectively operated by the ILT foundation under a joint scientific policy. The ILT resources have benefitted from the following recent major funding sources: CNRS-INSU, Observatoire de Paris and Universit\'{e} d'Orl\'{e}ans, France; BMFTR, MIWF-NRW, MPG, Germany; Science Foundation Ireland (SFI), Department of Business, Enterprise and Innovation (DBEI), Ireland; NWO, The Netherlands; The Science and Technology Facilities Council, UK.
This paper uses data obtained with the German LOFAR stations,
during station-owners time and ILT time allocated under project codes
LC0\_014, LC1\_048, LC2\_011, LC3\_029, LC4\_025, LT5\_001, LC9\_039, LT10\_014 and LT14\_006.
We made use of data from
the Effelsberg (DE601) LOFAR station funded by the Max-Planck-Gesellschaft;
the Unterweilenbach (DE602) LOFAR station funded by
the Max-Planck-Institut für Astrophysik, Garching;
the Tautenburg (DE603) LOFAR station funded by the State of Thuringia,
supported by the European Union (EFRE) and the BMFTR Verbundforschung
project D-LOFAR I (grant 05A08ST1);
the Potsdam (DE604) LOFAR station funded by the
Leibniz-Institut für Astrophysik (AIP), Potsdam;
the Jülich (DE605) LOFAR station supported by the
BMFTR Verbundforschung project D-LOFAR I (grant 05A08LJ1);
and the Norderstedt (DE609) LOFAR station funded by the
BMFTR Verbundforschung project D-LOFAR II (grant 05A11LJ1).
The observations of the German LOFAR stations
were carried out in stand-alone GLOW mode,
which is technically operated and supported by
the Max-Planck-Institut für Radioastronomie, the Forschungszentrum
Jülich and Bielefeld University. We acknowledge support and
operation of the GLOW network, computing and storage facilities by
the FZ-Jülich, the MPIfR and Bielefeld University and financial support
from BMFTR D-LOFAR III (grant 05A14PBA) and D-LOFAR IV (grant 05A17PBA),
and by the states of Nordrhein-Westfalia and Hamburg.
\end{acknowledgements}

\bibliography{main}
\bibliographystyle{aa}

\appendix
\section{Posterior probability distributions for the fitted parameters during MCMC}
\begin{figure}
    \centering
    \includegraphics[width=0.8\textwidth]{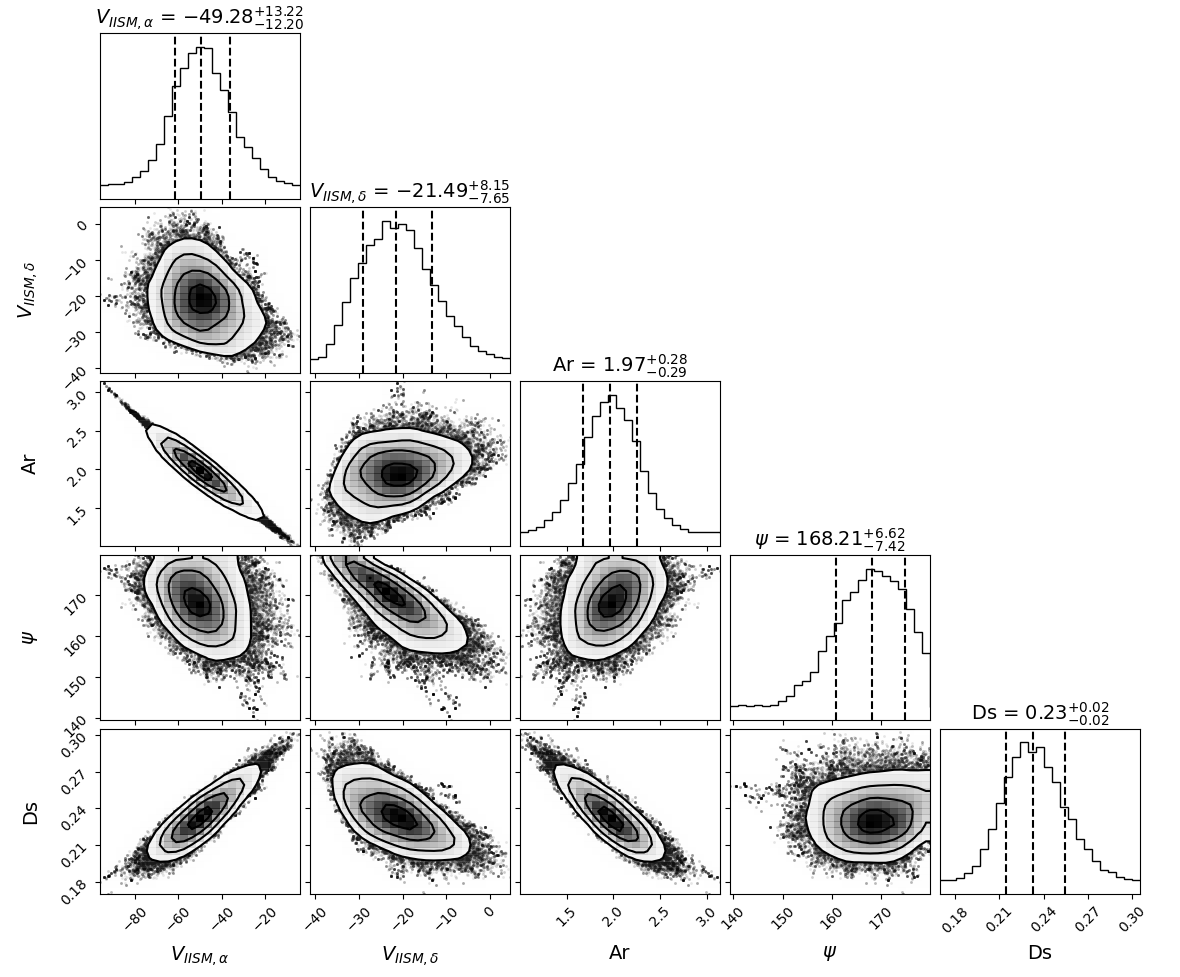}
    \caption{The posterior probability distributions of all fitted parameters. From left to right in each 1D histogram, three black dashed lines denote the 16\% fractional percentiles, the most likely values, and the 84\% fractional percentiles, respectively. The most likely values and the upper/lower errors are indicated at the top of the 1D histograms.}
    \label{fig:enter-mcmc}
\end{figure}

\end{document}